\documentclass[lettersize,journal]{IEEEtran}
\usepackage{amsmath,amsfonts}
\usepackage{array}
\usepackage[caption=false,font=normalsize,labelfont=sf,textfont=sf]{subfig}
\usepackage{textcomp}
\usepackage{stfloats}
\usepackage{url}
\usepackage{verbatim}
\usepackage{graphicx}
\usepackage{cite}
\usepackage{amsfonts,amssymb,amsmath,amsthm,amscd,stmaryrd,euscript,array,algorithm,algpseudocode}
\usepackage{makeidx} 
\usepackage{xcolor, multirow}
\usepackage[symbol]{footmisc}

\usepackage[margin=1in]{geometry} 
\usepackage{hyperref}

\newcommand{\ds}{\displaystyle}

\newtheorem{theorem}{\bf Theorem}[section]
\newtheorem{defi}[theorem]{\bf Definition}

\newtheorem{lemma}[theorem]{\bf Lemma}

\newtheorem{ex}[theorem]{\bf Example}
\newtheorem{rem}[theorem]{\bf Remark}
\newtheorem{prot}[theorem]{\bf Protocol}

\newtheorem{statement}{\bf Statement}[section]

\newenvironment{definition}{\begin{defi}\rmfamily\upshape}{\end{defi}}

\makeatletter
\newtheorem*{rep@theorem}{\rep@title}
\newcommand{\newreptheorem}[2]{%
\newenvironment{rep#1}[1]{%
 \def\rep@title{#2 \ref{##1}}%
 \begin{rep@theorem}}%
 {\end{rep@theorem}}}
\makeatother
\newreptheorem{theorem}{Theorem}

\newcommand{\Z}{\mathbb{Z}}

\newcommand{\Comm}{\textrm{COMM}}
\newcommand{\DPRF}{\textrm{DPRF}}
\newcommand{\TSSS}{\textrm{TSSS}}

\newcommand\Algphase[1]{%
\vspace*{-0.2\baselineskip}\Statex\hspace*{\dimexpr-\algorithmicindent-2pt\relax}\rule{8cm}{0.4pt}%
\Statex\hspace*{-\algorithmicindent}\textbf{#1}%
\vspace*{-.7\baselineskip}\Statex\hspace*{\dimexpr-\algorithmicindent-2pt\relax}\rule{8cm}{0.4pt}%
}

\usepackage{authblk}

\begin{document}

\title{Trustless Distributed Symmetric-key Encryption}

\author[1]{Florian Le Mouël\footnote{Work done as an intern at Kelvin Zero}}
\author[2]{Maxime Godon}
\author[3]{Renaud Brien}
\author[1]{Erwan Beurier}
\author[1]{Nora Boulahia-Cuppens}
\author[1]{Frédéric Cuppens}
\affil[1]{Polytechnique Montreal, \{florian.le-mouel, erwan.beurier, nora.boulahia-cuppens,frederic.cuppens\}@polymtl.ca}
\affil[2]{Kelvin Zero, maxime.godon@uqo.ca}
\affil[3]{Kelvin Zero, renaud.brien@gmail.com}



\maketitle

\begin{abstract}
	Threshold cryptography has gained momentum in the last decades as a mechanism to protect long term secret keys. Rather than having a single secret key, this allows to distribute the ability to perform a cryptographic operation such as signing or encrypting. Threshold cryptographic operations are shared among different parties such that a threshold number of them must participate in order to run the operation. This makes the job of an attacker strictly more difficult in the sense that they would have to corrupt at least a threshold number of parties to breach the security. Most works in this field focus on asymmetric-key schemes that allow threshold signing or decrypting.
	
	We focus on the symmetric-key setting, allowing both threshold encryption and threshold decryption. Previous work \cite{agrawal2018dise} relies on the presence of a trusted third party. Such a party may not exist in some use cases, and it represents a single point of failure. We propose to remove the requirement of a trusted third party by designing a dealer-free setup in which no entity can at any point obtain full knowledge of the secret keys.
	
	We implement a proof of concept of our construction in Python. We evaluate the proof of concept with timing metrics to compare to theoretical expectations and assess the cost in complexity of not relying on a trusted third party. While the setup phase suffers moderate additional cost, the encryption and decryption phases perform the same as the original algorithm.

\end{abstract}
\section{Introduction}
\label{introChapter}

	Symmetric-key encryption is a fundamental tool when it comes to the protection of sensitive data, but its usage comes with a critical task: protecting the encryption key. Knowledge of this key allows an adversary to trivially break the security of any data encrypted with it. Since the key as a whole gives such capability, one solution is to split it in pieces. This technique, better known as distributed encryption, stands out as one of the main alternatives to protect long-term secret keys, alongside encrypting the key itself and using secure hardware.

	While asymmetric distributed encryption has been well studied, we look into its less examined application to symmetric-key cryptography. Changing the mechanism to access the data, from a single key holder to a consensus of partial key holders, reinforces the security of the scheme. Compromising the data requires compromising more than one entity, making the attacker's job considerably more difficult. More specifically, we design a \emph{Threshold Symmetric-key Encryption} (TSE) scheme. A $k$-of-$n$ TSE scheme remains secure even if $k-1$ among $n$ participants are corrupted by an adversary.\\
	
	Distributed Symmetric-key Encryption (DiSE) \cite{agrawal2018dise} adapts security notions to distributed symmetric-key encryption and proposes a generic construction meeting these requirements. In their proposed instantiations however, the scheme requires a Trusted Third Party (TTP) to manipulate sensitive cryptographic data. This TTP introduces some limitations to the scheme. Firstly, it represents a single point of failure for confidentiality if it is compromised at the right time. Secondly, the responsibility of providing a TTP becomes a sensitive issue when instantiating DiSE between mutually non-trusting entities.
	
	The author's proposed application for DiSE is to distribute encryption capacities among servers belonging to the same entity (or to mutually trusting entities). This entity could be a company providing a long term encryption service. In this context, the entity can provide a TTP for the scheme.
	
	In a use case where each server belongs to a different entity, choosing an appropriate TTP could become a sensitive issue. A scheme without TTP is appropriate to distribute encryption capacities among entities without requiring that they trust each other. For example, a scheme without TTP would allow joint encryption of a file by a student, a school and a company such that at no point any of the three hold enough information to encrypt or decrypt the file without involving the others.\\
	
	Our proposed contributions are:
\begin{enumerate}
	\item A Distributed Key Generation (DKG) protocol for DiSE's strongly secure DDH-based Distributed Pseudo-Random Function \cite{agrawal2018dise}.
	\item Using this DKG protocol to bring DiSE \cite{agrawal2018dise} into the trustless context.
	\item A proof of our encryption protocol's security in the Random Oracle Model assuming a static corruption model and the hardness of the Decisional Diffie-Hellman (DDH) problem.
	\item An instantiation of Trustless Distributed Symmetric-key Encryption (TDiSE) guaranteeing confidentiality, strong-correctness and strong-authenticity without relying on a TTP.
	\item A comparison of timing complexity between our scheme and DiSE supported by measurements on implementations of both schemes.
\end{enumerate}

	The sections of this article are organized as follows. In Section \ref{relatedWorksChapter}, we discuss existing results on related subjects including secret sharing, DPRFs, distributed asymmetric-key cryptography and distributed symmetric-key cryptography. Section \ref{backgrndChapter} covers technical prerequisites for our scheme, including details on the construction proposed in DiSE \cite{agrawal2018dise}. Section \ref{mainChapter} details the construction of our protocol and Section \ref{implementationChapter} covers its implementation. Finally, we explain how to implement proactive security with our scheme in Section \ref{proactiveChapter}.

\section{Related works}
\label{relatedWorksChapter}

	\subsubsection{Secret Sharing Schemes} A secret sharing scheme allows to share a secret value between 2 or more entities without revealing the secret to individual entities. In 1979, Shamir published a seminal paper \cite{shamir1979share} that gave birth to the domain of threshold cryptography. While a secret sharing scheme does not meet the required properties for an encryption scheme, it is a useful primitive to design distributed encryption schemes such as ours.
	
	The same year, Blakley published another threshold secret sharing scheme \cite{blakley1979safeguarding}. However, the keys in \cite{shamir1979share} have constant size for a given security parameter, which is not the case for \cite{blakley1979safeguarding}. Later works like \cite{ito1989secret} proposed constructions for secret sharing schemes implementing access structures more general than threshold structures. \cite{beimel2011secret} provides more examples and discussions on secret sharing schemes and their applications.
	
	Our contributions have close ties with Verifiable Secret-Sharing (VSS) as coined by \cite{chor1985verifiable}. A VSS scheme is a secret sharing scheme that allows protection against the other parties, including the distributor, by ensuring that the secret shares allow reconstruction of a consistent secret. VSS opens the possibility to implement Distributed Key Generation (DKG) protocols to replace the TTP. The resulting scheme is called a joint random secret sharing. A popular example was proposed by Feldman in 1987 \cite{feldman1987practical} and later improved by Pedersen \cite{pedersen1991non} and then Gennaro, Jarecki, Krawczyk and Rabin \cite{gennaro2007secure}. Benaloh's scheme \cite{benaloh1986secret} is also often cited.\\
	
	\subsubsection{Distributed Pseudo-Random Functions (DPRF)} Shamir's Secret Sharing Scheme \cite{shamir1979share} (SSSS) is the basis for a number of DPRFs. DPRFs provide a useful primitive for the design of a distributed symmetric-key encryption scheme \cite{naor1999distributed}.
	
	Considering that standard assumptions such as DDH will likely be broken in a few decades by the emergence of quantum computers, a lot of the modern works in this field shifted to Learning With Error (LWE) based constructions \cite{boneh2018threshold,boneh2013key,kim2020key,libert2021adaptively}. These could provide post quantum security, but come with a higher computational cost than other DPRF constructions. Other works like \cite{darivandpour2021secure} restrict their schemes to the two-party setting, where they achieve obliviousness for their DPRF, using xor-PRFs or any oblivious PRF.
	
	We are particularly interested in Distributed Verifiable Random Functions (DVRF), which instantiate a strongly-secure DPRF as defined later in Definition \ref{DPRFsec}. \cite{galindo2021fully} proposes two Distributed Verifiable Random Functions: one based on the DDH problem and the other based on pairings. Their constructions, like ours, do not require the intervention of a trusted third party, as they exploit the DKG protocol introduced in \cite{gennaro2007secure}. Our construction, while similar, does not ensure identifiable aborts, nor the robustness that follows in \cite{gennaro2007secure}. In the use cases considered for TDiSE, excluding a party from generation would lead to an undesired result where the excluded party has no guarantee over the randomness of the shared secret. In exchange, our DKG benefits from lighter amounts of computation.\\
	
	\subsubsection{Distributed Asymmetric-key Cryptography} Asymmetric-key cryptography is complementary to symmetric-key cryptography, therefore an overview of distributed asymmetric-key cryptography will provide both relevant problems and relevant primitives for distributed symmetric-key cryptography.
	
	Distributed signature schemes distribute the ability to sign data among several participants. Using a public key, an entity can verify that a signature was produced using the corresponding set of private keys. While standard schemes have been thresholdized more than 20 years ago \cite{gennaro1996robust}, a majority of the recent works focus on Elliptic Curve Digital Signature Algorithm (ECDSA)  \cite{damgaard2022fast,canetti2020uc}, or on alternative elliptic curve based signature schemes like Edwards-curve Digital Signature Algorithm (EdDSA) \cite{bonte2021thresholdizing}. Other techniques include the use of lattices \cite{pilaram2021efficient} and isogenies \cite{cozzo2020sashimi} to tackle the growing interest in quantum secure cryptography.
	
	Distributed public key encryption schemes distribute the ability to decrypt data. Using a public key, an entity can encrypt data such that it can only be decrypted with the corresponding set of private keys. A lot of work is invested in thresholdizing existing public-key encryption schemes such as the Paillier cryptosystem \cite{devevey2022rational} and Rivest-Shamir-Adleman (RSA) \cite{buldas2017server} or using established cryptographic primitives like LWE \cite{alborch2022r,boneh2018threshold} or bilinear pairings \cite{gao2018efficient} to produce threshold public-key encryption schemes. We also find a variety of proposed functionalities like identity-based encryption \cite{gao2018efficient}, re-splittability of keys \cite{ohata2018more} or searchable encryption \cite{miao2020threshold}.\\
	
	\subsubsection{Distributed Symmetric-key Cryptography} Distributed symmetric-key cryptography achieves different goals. All participants in a distributed symmetric-key encryption scheme own keys from a set of secret keys. This set of secret keys allows both encryption and decryption. In opposition to distributed asymmetric-key encryption schemes, the ability to encrypt data is not public.
	
	To the best of our knowledge, the only existing TSE schemes at the time of writing are DiSE \cite{agrawal2018dise}, Amorthized Threshold Symmetric-key Encryption (ATSE) \cite{christodorescu2021amortized} and Hierarchical (Threshold) Symmetric-key Encryption (HiSE) \cite{dey2024hise} based on ATSE. The concept of TSE is new considering that DiSE first introduced formal definitions of it. DiSE also benefits from a few upgrades to security properties, namely \cite{wang2020robust}, \cite{sinha2024efficient} and \cite{mukherjee2020adaptively}.
 
	\cite{wang2020robust} proposes a method to add strong correctness to DiSE's PRF-based construction. They do so by exploiting the redundancy of keys in the underlying DPRF, based on any PRF, originally proposed by \cite{naor1999distributed}. This implementation however is not suitable for practical scalable  implementations, since the size of keys scales exponentially with the threshold parameter.
    \cite{sinha2024efficient} proposes an efficient instantiation of DiSE relying on quantum-safe assumptions, taking the first step towards post-quantum distributed symmetric-key cryptography. Their scheme uses a secure PRF based on the learning with rounding problem, which is proven secure in the usual pre-quantum random oracle model.
	\cite{mukherjee2020adaptively} makes DiSE secure against a dynamic adversary. Their construction is therefore secure, even if the adversary can corrupt new servers after the setup phase.
 
	Recently, Distributed Authenticated Encryption (DiAE) \cite{duc2022diae} pointed at flaws in the implementations proposed by DiSE that therefore fail to ensure strong correctness. DiAE also proposes fixes to correct the aforementioned flaws. ParaDiSE \cite{agrawal2022paradise} identifies shortcomings in the confidentiality and integrity definitions proposed by DiSE and proposes a new security definition that combines both. The strongly secure construction of DiSE can be proven secure under a modified version of their security definition, as discussed in the appendix of ParaDiSE.
	
	We derive our construction from DiSE's DDH-based construction which is strongly secure against a static adversary by modifying the setup phase. We take into account the implementation flaws \cite{duc2022diae} pointed at and incorporate them in our own implementations.
	
\section{Background}
\label{backgrndChapter}

\subsection{Notations}

	Let $\mathbb{N}$ denote the set of positive integers and $\mathbb{Z}$ denote the set of all integers. For any prime integer $q \in \mathbb{N}$, we use $\mathbb{Z}_q$ to denote the finite field of integers modulus $q$. For the rest of this paper we use $\lambda$ to denote the security parameter, $p$ and $q$ to denote large primes such that $q$ is the largest prime factor of $p-1$ and is a $\lambda$-bit value. The total number of parties will be denoted by $n$ and the threshold parameter will be denoted by $k$. For DPRFs, the threshold parameter is the minimum number of parties required to compute the output. For TSE schemes, it represents the minimum number of parties required to either encrypt or decrypt a message. When considering an adversary, $t$ will denote the number of parties corrupted by the adversary.
	
	For $d\in \mathbb{N}$, we denote by $\Z_q^d[X]$ the linear space of polynomials with coefficients in $\Z_q$ of degree at most $d$. For $m \in \mathbb{N}$, we use $\left\llbracket m\right\rrbracket$ to denote the set of integers $\{1, 2, ..., m\}$. Each party will be labeled with a different integer $i\in\left\llbracket n\right\rrbracket$. We use $a||b$ to denote the concatenation of two strings $a$ and $b$. The xor operation between two byte strings $a$ and $b$ is denoted $a \oplus b$.

\subsection{Prerequisites}

	\subsubsection{Commitment schemes} A commitment scheme allows a party to generate data bound to a secret value without giving further information about the secret value. The commitment can be unveiled later with the guarantee that the value is the same as committed.

	\begin{definition}
		\textbf{(Commitment scheme)} Let COMM consist in three algorithms: \\$\langle setup_\Comm, commit_\Comm, verify_\Comm\rangle$ such that:
		\begin{itemize}
			\item $setup_\Comm(\lambda)$ outputs public parameters $pp_\Comm$ for the security parameter\footnote{The length of messages that can be committed to is determined by $\lambda$. For longer messages, one should either instantiate a commitment scheme with a higher $\lambda$ value or hash their message to a string of appropriate length.} $\lambda$
			\item $commit_\Comm(pp_\Comm, m, r)$ outputs a commitment $\alpha$ to the message $m$ with randomness $r$
			\item $verify_\Comm(pp_\Comm, m, r, \alpha)$ outputs true if \\$\alpha = commit_\Comm(pp_\Comm, m, r)$, false otherwise
		\end{itemize}
		COMM is a commitment scheme if, for all public parameters $pp_\Comm$, it satisfies the following two properties:

		\emph{Hiding:} For any messages $m_0, m_1$, given $\alpha = commit_\Comm(pp_\Comm, m, r)$, with $m\in \{ m_0, m_1\}$ and $r$ a randomly sampled value unknown to the adversary, an adversary can guess $m$ with a negligible advantage

		\emph{Binding:} For any message $m$ and randomness $r$, an adversary can produce $m', r'$ such that $commit_\Comm(pp_\Comm, m', r') = commit_\Comm(pp_\Comm, m, r)$ with negligible probability
	\end{definition}

	Those define a computationally hiding and binding commitment scheme. If the probability of breaking hiding (resp. binding) is $\frac{1}{2}$ (resp. $0$), we say that the commitment has \emph{perfect hiding (resp. binding)}.\\
	
	\paragraph{Instantiations} We will be interested in two commitment schemes in our instantiation: the Pedersen commitment scheme and the hash-based commitment scheme.
	
	The Pedersen commitment scheme is a perfectly hiding and computationally binding commitment scheme. It requires two generators $g, h$ of the same group of order $q$ such that the discrete logarithm of $h$ relative to $g$ is unknown to any party that runs $commit_{Pedersen}$. Knowledge of that discrete logarithm would allow to break the binding property. While there are many possible ways to generate such a pair, we require that $setup_{Pedersen}$ does not require a TTP. We discuss how a party should proceed in Section \ref{implementationChapter}. In order to commit to a message $m$ using the Pedersen commitment scheme, a party samples $r\in\Z_q$ and computes: $$\gamma = g^m*h^r$$ This party must store $r$ privately to be able to unveil the committed message. We will refer to the algorithms of the Pedersen commitment scheme as $ setup_{Pedersen}$, $commit_{Pedersen}$ and $verify_{Pedersen}$.
	
	The hash-based commitment scheme is a computationally hiding and binding commitment scheme in the random oracle model. It uses a hash function $H_{comm}$ which outputs byte strings $3$ times longer that the committed message to guarantee security requirements. Several variants exist depending on the choice for $H_{comm}$. The hash-based commitment scheme does not require a setup. A party that wishes to commit to a message $m$ samples $r$ of size equal to $m$ and computes: $$\alpha = H_{comm}(r||m)$$ We will refer to the algorithms of the hash-based commitment scheme as \\$setup_{hash-based}$, $commit_{hash-based}$ and $verify_{hash-based}$.

	\subsubsection{Threshold secret-sharing schemes} A \emph{$k$-of-$n$ threshold secret-sharing scheme} is a scheme that allows sharing of a secret among $n$ parties such that any coalition of $k$ or more of them can recover the secret, and any coalition of $k-1$ or less parties learn nothing about the secret. We will use a similar definition to the one proposed by \cite{beimel2011secret}. Since we are only interested in threshold constructions, we will restrict our definition to the threshold case.
	
	\begin{definition}
		\textbf{(Threshold secret-sharing scheme).}\footnote{This definition could easily be adapted for non threshold constructions by changing the conditions in consistency and privacy. In a more general definition, one could define a list of all authorized subsets that would be allowed to reconstruct the secret. Consistency would hold for all authorized subsets and Privacy for all other subsets.} Let $Participants$ be a set of $n$ parties. Let TSSS consist in two algorithms $\langle generate_\TSSS, combine_\TSSS \rangle$ such that, for a secret $s$, $generate_\TSSS(\lambda, k, n, s,  Participants)$ outputs public parameters $pp_\TSSS$ and the set of keys $SK=\{sk_1, sk_2,..., sk_n\}$ such that each key $sk_i$ for $i\in [1,n]$ is given to party $p_i\in Participants$.
		
		TSSS is a $k$-of-$n$ threshold secret sharing scheme if the following two requirements hold:
	
	\emph{Consistency} For all subsets $K\subseteq Participants$ such that $K$ comprises $k$ or more parties: $$combine_\TSSS(pp_\TSSS, \{sk_j\}_{j\in K}) = s$$
	
	\emph{Privacy} For all subsets $T\subset Participants$ such that $T$ comprises at most $k-1$ parties, for all pairs of secrets $s, s'$ and all possible secret key subset $SK_T = \{sk_j\}_{j\in T}$, a probabilistic polynomial time adversary can guess whether $SK_T$ was generated with $generate_\TSSS(\lambda, k, n, s, Participants)$ or \\$generate_\TSSS(\lambda, k, n, s' Participants)$ with a negligible advantage.
	
	This defines computational privacy. If the probability of breaking privacy is $\frac{1}{2}$, we say that the scheme has \emph{perfect privacy}. 
	\end{definition}
	
	\paragraph{Instantiation} In this work, we use Shamir's Secret-Sharing Scheme (SSSS) \cite{shamir1979share}. SSSS (Algorithm \ref{SSSS}) splits a secret value $s\in \llbracket 0, q-1\rrbracket$ among $n$ parties by distributing points on a degree $k-1$ polynomial. We reconstruct the secret by using Lagrange interpolation on $k$ or more of these points, which yields the polynomial.
	
	\subsubsection{Distributed Pseudo-Random Functions} A DPRF distributes the evaluation of a secret PRF. In order to evaluate a DPRF, a party must obtain the help from an authorized set of participants. To be more specific, since we are interested in threshold constructions, we will discuss threshold DPRFs. A threshold DPRF is distributed among $n$ parties such that any coalition of $k$ or more of them can compute the function on an input they agreed on. The evaluation of the function on a new input is pseudo-random to any coalition of at most $k-1$ parties.
	We will refer to the definition proposed in \cite{agrawal2018dise} for a DPRF assuming perfect consistency:
	
	\begin{definition}
	\label{DPRF}
		\textbf{(Threshold Distributed Pseudo-Random Function).} Let $Participants$ be a set of $n$ parties. Let DPRF consist in three algorithms \\$\langle generate_\DPRF, evaluate_\DPRF, combine_\DPRF \rangle$ such that:
		\begin{itemize}
			\item $generate_\DPRF(\lambda, k, n, Participants)$ outputs public parameters $pp_\DPRF$ for the security parameter $\lambda$ and $SK=\{sk_1, sk_2, ..., sk_n\}$ such that each key $sk_i$ for $i\in \llbracket n \rrbracket$ is only known to party $p_i\in Participants$ and is never known any $p\in Participants$ for $p \neq p_i$.
			\item $evaluate_\DPRF(pp_\DPRF, sk_i, x)$ outputs a pseudo-random partial evaluation $z_i$
			\item $combine_\DPRF(pp_\DPRF, \{z_i\}_{i\in S})$ combines the partial evaluations from $S\subseteq Participants$ to obtain the evaluation $z$ of the DPRF.
	\end{itemize}
		DPRF is a threshold distributed pseudo-random function if the \emph{consistency} requirement holds:
			
			\emph{(Consistency)} For all subsets $K, K'\subseteq Participants$ such that $K$ and $K'$ both comprise $k$ or more parties and for all input $x$: $combine_\DPRF(pp_\Sigma, \{z_i\}_{i\in K}) = combine_\DPRF(pp_\Sigma, \{z_i\}_{i\in K'})$ where $z_i = evaluate_\DPRF(sk_i, x)$
	\end{definition}

	\begin{definition}
	\label{DPRFsec}
	\textbf{(Security of a DPRF).} A threshold DPRF will be considered \emph{secure} if the pseudo-randomness property holds against an adversary that has corrupted at most $k-1$ parties. 
	
	\emph{(Pseudo-randomness)} The \emph{pseudo-randomness} property for a DPRF states that an adversary that has access to an oracle can guess a new evaluation of the DPRF with probability at most $\frac{1}{2^\lambda}$.
	
	A threshold DPRF will be considered \emph{strongly-secure} if both the pseudo-randomness and correctness properties hold against an adversary that has corrupted at most $k-1$ parties.
	
	\emph{(Correctness)} The \emph{correctness} property for a DPRF states that an adversary that has access to an oracle can tamper with the output of an evaluation without being detected with negligible probability at best.
	
	A \emph{trustless\footnote{While this is effectively implemented using a dealer-free DPRF, we use the term trustless to emphasize that such a function is meant to be used in the context of mutually untrusting parties.} DPRF} is a DPRF such that each party's secret key is known only to them and no party ever holds enough information to compute the DPRF on its own. Such a scheme cannot use a TTP to generate and distribute the keys.

	We refer to \cite{agrawal2018dise} for formal definitions of pseudo-randomness and correctness.
	\end{definition}
	
	\paragraph{Instantiations} One example of secure DPRF construction combining the DDH problem and SSSS \cite{shamir1979share} is proposed by \cite{naor1999distributed}. The pseudo-code is specified in Algorithm \ref{NPR}.

	This construction can be made strongly secure by adding Non-Interactive Zero Knowledge proofs to the output of $evaluate_{NPR}$. Specifically, it uses a proof system similar to the Chaum-Pedersen protocol for Diffie-Hellman-triples \cite{chaum1993wallet}. The generation of secret keys starts the same way as $generate_{NPR}$ but additionally outputs a Pedersen commitment $\gamma_j$ to each secret key $sk_j$.
	
	The evaluation requires a new hash function $H_{chal}$ with output space $\mathbb{Z}_q$ to make the proofs non-interactive thanks to the Fiat-Shamir transform. $H_{chal}$ will be used by the provers (the parties that run $evaluate_{SSNPR}$) to obtain a pseudo-random challenge.
	
	Note that the outputs of $evaluate_{SSNPR}$ include the output of $evaluate_{NPR}$ with additional data to check the computation. \\$combine_{SSNPR}$ uses this additional data for verification purposes only. The output of $combine_{SSNPR}$ is therefore the same as \\$combine_{NPR}$.
	
	The pseudo-code is given in Algorithm \ref{SSNPR}.
	
	In Section \ref{mainChapter}, we discuss how to turn $generate_{SSNPR}$ into a dealer-free $generate$ algorithm.

\subsection{Distributed Symmetric-key Encryption (DiSE)}

	DiSE \cite{agrawal2018dise} is the first formal approach to distributed symmetric-key encryption. It formalizes appropriate security definitions for authenticated encryption in the distributed setting and proposes an encryption scheme based on any DPRF and implementations using \cite{naor1999distributed}.
	The novel security definitions include (strong)-correctness, message privacy and (strong)-authenticity. The interested reader may find these definitions in \cite{agrawal2018dise}.\\
	
\begin{definition}
\label{TSESecurity}
	\textbf{(Security of a TSE scheme).} A TSE scheme is called \emph{secure} if correctness, message privacy and authenticity requirements hold. It will be considered \emph{strongly-secure} if strong-correctness, message privacy and strong-authenticity hold.
\end{definition}

	DiSE's proposed construction composes any DPRF with a commitment scheme in order to add authenticity and correctness to the scheme (Algorithm \ref{DiSE}). If the DPRF is secure (resp. strongly-secure), then the resulting TSE is a secure (resp. strongly-secure) TSE.

	In this article we are particularly interested in a strongly-secure instantiation of the previous construction, using $SSNPR$ as a DPRF and the $hash-based$ commitment scheme as components for DiSE's blueprint (Algorithm \ref{DiSE}).

\section{Trustless Distributed Symmetric-key Encryption}
\label{mainChapter}

	In Algorithm \ref{DiSE}, DiSE requires a TTP to run the setup algorithm, as it manipulates the secret keys of the participants. On the one hand, compromising this TTP before setup allows an attacker to access all the secret data of the scheme and to distribute invalid keys. On the other hand, removing the need of a TTP also removes the task of assessing this responsibility. The following section explains the design of a Trustless Distributed Symmetric-key Encryption (TDiSE).

\subsection{Setting overview}

	The environment includes 3 types of parties.
	
		\textbf{Participants} are the owners of the secret keys. They collaborate in order to run encryption and decryption operations. All the $n$ participants that will potentially take part in encryption and decryption operations must be online in order complete the setup phase, where they generate the secret keys. However, only $k$ of them have to take part in encryption and decryption operations; the rest of the participants may be offline as those operations take place.
		
		A \textbf{public storage} is needed to store the encrypted data and public parameters and make them available to participants at all times. The public storage does not store any sensitive data.
		
		An \textbf{untrusted third party} is responsible for initiating the setup phase. This role could be assumed by any participant, or an entity external to the scheme. The untrusted third party is not essential to the scheme, but it is useful to initiate the setup and keep track of the setup's advancement. This party does not manipulate sensitive data and could even be malicious without compromising the scheme. In other schemes such as DiSE \cite{agrawal2018dise}, TTPs are only used as part of the setup phase. As such, only the setup phase needs to be modified. Since the state of the game is identical at the end of the setup phase in our proposed scheme and in DiSE, we defer to their proof of security for the encryption and decryption phases.

\subsection{Removing the trusted third party}

	SSSS is a linear secret-sharing scheme, meaning that given two secrets $s_1, s_2$, the secret keys obtained by running $generate_{SSSS}(q(\lambda), k, n, Participants, s_1)$ and $generate_{SSSS}(q(\lambda), k, n, Participants, s_2)$ can be linearly combined to obtain a secret-sharing of the same linear combination of $s_1$ and $s_2$. This property of SSSS \cite{shamir1979share} is easily extended to an arbitrary number of key sets.

\begin{statement}
	\textbf{(Linearity of SSSS).}\\
	For $i\in \llbracket n\rrbracket$ and $s_i\in \mathbb{Z}_q$, let $\{x_j\}_{j\in \llbracket n \rrbracket}\in (\Z_q)^n$ and $\{sk_{i,j}\}_{j\in Participants}$ be the set of keys generated by $generate_{SSSS}(q(\lambda), k, n, \{x_j\}_{j\in \llbracket n \rrbracket}, Participants, s_i)$ then for any $K\subseteq Participants$ of size at least $k$: $$\ds combine_{SSSS}\left(q, k, n, \left\{\sum_{i=1}^n sk_{i,j}\right\}_{j\in K}\right) = \sum_{i=1}^n s_i$$
\end{statement}

	It follows that since NPR, as presented in Algorithm \ref{NPR}, uses a set of SSSS keys, $\ds\left\{\sum_{i=1}^n sk_{i,j}\right\}_{j\in Participants}$ is also a valid set of keys for NPR. NPR is derived from a key-homomorphic PRF. We can therefore deduce the following property about NPR \cite{naor1999distributed}:
\begin{statement}
\label{homomorphic}
	\textbf{(Key homomorphism of NPR).}\\
	For $i\in \llbracket n\rrbracket$, let $\{x_j\}_{j\in \llbracket n \rrbracket}\in (\Z_q)^n$ and $\{sk_{i,j}\}_{j\in Participants}$ be the set of keys generated by $generate_{NPR}(q(\lambda), p(\lambda), k, n, \{x_j\}_{j\in \llbracket n \rrbracket}, Participants)$.
\begin{itemize}
	\item For $j\in Participants$:
    $$\ds\prod_{i=1}^n evaluate_{NPR}(p, q, sk_{i,j}, x) = $$
    $$evaluate_{NPR}\left(p, q, \sum_{i=1}^n sk_{i,j}, x\right)$$
	\item For $j\in Participants$, let $z_{i,j} = evaluate_{NPR}(p, q, sk_{i,j}, x)$, for any $K\subseteq Participants$ of size at least $k$:
    $$\ds\prod_{i=1}^n combine_{NPR}(p, q, \{z_{i,j}\}_{j\in K}) = $$
    $$combine_{NPR}\left(p, q, \left\{\prod_{i=1}^n z_{i,j}\right\}_{j\in K}\right)$$
\end{itemize}
\end{statement}

\subsection{The protocol}

	In order to setup the scheme without a TTP, each of the $n$ participants must be aware of the others and have secure one to one communication channels with them.\footnote{The participants need secure end to end communication channels before getting involved in the scheme. The channels must ensure confidentiality and integrity. An adversary eavesdropping the communications of the setup phase should not gain information on the partial keys.}
	
	The untrusted third party prompts every participant to run Algorithm \ref{AlgwithoutTTP}. In order to avoid malicious tampering in the key generation, after commitments to the keys have been published, the untrusted third party prompts each participant  at line 7 of Algorithm \ref{AlgwithoutTTP} to run a test encryption to verify that the result is a $k-of-n$ threshold DPRF. The test's input message is generated through a hash function $H_{test}$ to bind the test value to the public parameters and avoid malicious test values. If any of the assertions at the end of Algorithm \ref{AlgwithoutTTP} fails, the setup is aborted and must be started from the beginning.
	
	This setup algorithm uses commitments, therefore these must also be initiated first. For the hash-based commitment scheme, the public hash function used can be the same across all instantiation, therefore it does not need to be setup. We discuss how an untrusted party can setup the Pedersen commitment scheme in Section \ref{implementationChapter}. Once the commitment schemes are setup, each participant $i$ must run Algorithm \ref{AlgwithoutTTP}.

\begin{algorithm}[h]
\caption{$Setup_{TDiSE}$ run by participant $i$}
\label{AlgwithoutTTP}
\fontsize{8pt}{8pt}\selectfont
\begin{algorithmic}[1]
	\Require $p(\lambda), q(\lambda), k, n, \{x_j\}_{j\in \llbracket n \rrbracket}, Participants$
	\Ensure $p, q$
	
	\hspace{-4em} \textbf{Private output:} $SK$
	\State Run $generate_{NPR}(p(\lambda), q(\lambda), k, n, \{x_j\}_{j\in \llbracket n \rrbracket}, Participants)$ to distribute secret keys $(sk_{i,1}, ..., sk_{i,n})$
	\For{$j\in Participants$}
		\State Receive $sk_{j,i}$ from participant $j$
	\EndFor
	\State Compute $sk_i = \sum\limits_{j\in\llbracket n\rrbracket} sk_{j, i}$
	\State Publish $\gamma_i = commit_{Pedersen}(sk_i)$
	\State Wait for every $j\in Participants$ to publish a commitment
	\State Generate randomness $\rho$
	\State Generate message $m = H_{test}(p, q, \lambda, k, n, Participants, \{\gamma_j\}_{j\in Participants})$
	\State Compute commitment $\alpha = commit_{hash-based}(m, \rho)$
	\For{ $j \in Participants$  }
		\State Send $i || \alpha$ to participant $j$
		\State Obtain $z_j = evaluate_{SSNPR}(p, q, sk_j, i || \alpha)$ from participant $j$
	\EndFor
	\State Assert that for $K\subset Participants$ of size exactly $k$, 
	
	\hspace{-2em} $combine_{SSNPR}(p, q, \{z_{i,j}\}_{j\in K}) = combine_{SSNPR}(p, q, \{z_{i,j}\}_{j\in Participants})$
	\State Assert that for $T\subset Participants$ of size exactly $k-1$, 
	
	\hspace{-2em} $combine_{SSNPR}(p, q, \{z_{i,j}\}_{j\in T}) \neq combine_{SSNPR}(p, q, \{z_{i,j}\}_{j\in Participants})$
	\State Assert that $combine_{SSNPR}(p, q, \{z_{i,j}\}_{j\in Participants}) \neq 1$
	\State \textbf{Return} $p, q$
\end{algorithmic}
\end{algorithm}

At the end of the setup, each participant has generated a set of keys for an NPR instantiation. Each participant then knows one key from each of these sets. As pointed out by Statement \ref{homomorphic}, when the participants sum their keys, they now share a set of keys for a new instantiation of NPR. However, at no point in the protocol does any participant hold enough information to reconstruct the secret.

\subsection{Proof of security}

	\subsubsection{Adversary model}	We consider a static adversary model, where an adversary could compromise a number $t$ of participants before setup that remains constant for the rest of the scheme. Such an adversary has therefore access to all the data manipulated by those $t$ participants, in particular the partial evaluations they compute for other participants. The adversary's goal is to break either strong-correctness, message privacy or strong-authenticity. We analyze the security of our scheme in the random oracle model and assuming the hardness of the DDH problem. 
	
\begin{theorem}
	Our TSE scheme is strongly-secure, as defined in Definition \ref{TSESecurity}, under the conditions that the participants communicate through secure end-to-end communication channels during setup and that $t < k$.
\end{theorem}

\begin{proof}
	The authors of DiSE \cite{agrawal2018dise} state that their generic construction is strongly-secure if instantiated with a strongly secure DPRF. The modifications we bring only concern the setup of the DDH-based DPRF from \cite{naor1999distributed}, therefore we will prove that our scheme is a strongly-secure TSE scheme by proving that the DPRF resulting of our modifications is strongly-secure. As stated in Definition \ref{DPRFsec}, our underlying DPRF is a strongly-secure threshold DPRF if the following three requirements hold: consistency, pseudo-randomness and correctness. In this section we will refer to the 3 algorithms of our DPRF simply as $\langle generate, evaluate, combine \rangle$.

	Throughout this proof, coefficients of polynomials are in $\mathbb{Z}_q$ with $q$ a $\lambda$-bit prime. Let $Participants$ be a set of $n$ parties, such that each party successfully completed Algorithm \ref{AlgwithoutTTP}.\\
	
	\textbf{Consistency.} For $i\in \llbracket n\rrbracket$, let $\{x_j\}_{j\in \llbracket n \rrbracket}\in (\Z_q)^n$ and $\{sk_{i,j}\}_{j\in Participants}$ be the set of keys generated by \\$generate_{NPR}(p(\lambda), q(\lambda), k, n, \{x_j\}_{j\in \llbracket n \rrbracket}, Participants)$. For $j\in Participants$, let $z_{i,j} = evaluate_{NPR}(p, q, sk_{i,j}, x)$.
	
	As seen in Algorithm \ref{AlgwithoutTTP}, for $j\in Participants$, $sk_j = \sum\limits_{i\in\llbracket n\rrbracket} sk_{i,j}$. Therefore, according to Statement \ref{homomorphic}:
	\begin{align*}
	\ds
	z_j & = evaluate_{NPR}(p, q, sk_j, x)\\
	& = evaluate_{NPR}\left(p, q, \sum_{i=1}^n sk_{i,j}, x\right)\\
	& = \prod_{i=1}^n evaluate_{NPR}(p, q, sk_{i,j}, x)\\
	& = \prod_{i=1}^n z_{i, j}
	\end{align*}
	 Let $K, K'\subseteq Participants$ such that both $K$ and $K'$ comprise $k$ or more parties. According to Statement \ref{homomorphic}:
	 \begin{align*}
	 \ds
	 combine(p, q, \{z_j\}_{j\in K})& = combine\left(p, q, \prod_{i=1}^n\{z_{i,j}\}_{j\in K}\right)\\
	 & = \prod_{i=1}^n combine(p, q, \{z_{i,j}\}_{j\in K})
	 \end{align*}

	Since $combine = combine_{SSNPR}$ and $SSNPR$ is a strongly-secure DPRF, the consistency property ensures that: $\forall i\in \llbracket n\rrbracket \text{, } combine(p, q, \{z_{i,j}\}_{j\in K}) = combine(p, q, \{z_{i,j}\}_{j\in K'})$
	
	Therefore: 
	\begin{align*}
	\ds
	combine(p, q, \{z_j\}_{j\in K})& = \prod_{i=1}^n combine(p, q, \{z_{i,j}\}_{j\in K})\\
	& = \prod_{i=1}^n combine(p, q, \{z_{i,j}\}_{j\in K'})\\
	& = combine(p, q, \{z_j\}_{j\in K'})
	\end{align*}
	
	\textbf{Pseudo-randomness.} In order to prove the pseudo-randomness property, we will first prove properties on SSSS key generation, before showing how they extend to our DPRF construction. We will need the following property:
	
	\begin{statement}
	\label{Uniformity}
		The sum $P + R$ of any polynomial $P\in \Z_q^{k-1}[X]$ with a uniformly sampled degree $k-1$ polynomial $R\in \Z_q^{k-1}[X]$ is a uniformly random polynomial in $\Z_q^{k-1}[X]$.
	\end{statement}
	
	Based on Statement \ref{Uniformity}, we prove two lemmas:
	
\begin{lemma}
\label{randomness}
	If at least one of the $n$ participants is honest, the shared SSSS polynomial resulting of the setup is uniformly random in $\Z_q^{k-1}[X]$.
\end{lemma}

\begin{proof}[Proof of Lemma \ref{randomness}]
	We assume that at least one participant in the setup is honest. Let $R\in \Z_q^{k-1}[X]$ be the polynomial sampled by this honest participant and $P$ be the sum of the polynomials sampled by all other participants. The shared SSSS polynomial will be $P+R$. Since polynomial addition is commutative, the order in which other polynomials are summed does not matter.
	
	\emph{Case 1:} If $P\in \Z_q^{k-1}[X]$, then $P + R$ is a uniformly random polynomial in $\Z_q^{k-1}[X]$ based on Statement \ref{Uniformity}.
	
	\emph{Case 2:} If $P$ is of degree $d > k-1$, then $P + R$ is also of degree $d$. In this case running Lagrange interpolation with $k$ participants will output a polynomial of degree at most $k-1$ which is therefore different from $P + R$. In order to detect such a malicious generation of $P$, the honest user will ask for the $n$ points and compare a Lagrange interpolation using $k$ points and a Lagrange interpolation using $n$ points. The interpolation with $k$ points yields a polynomial in $\Z_q^{k-1}[X]$, whereas the interpolation using $n$ points yields a degree $min(d, n-1)$ polynomial. If the honest user obtains two different polynomials, it means that at least one polynomial in $P$ was maliciously generated and they can abort the setup. Of course, using this technique would reveal the secret to the honest user in the case where $P$ was generated honestly, which renders the secret-sharing unusable.
	
	\emph{Remark:} In the corner case where $k=n$, interpolation yields a degree $n-1 = k-1$ polynomial which will remain consistent in the rest of the scheme. In that case, although $d > k-1$, the participants still share a uniformly random polynomial in $\Z_q^{k-1}[X]$. This is due to the fact that the same set of $n$ keys will be used for every interpolation.
	
	Revealing the secret can be avoided as we are not using the secret-sharing itself in our scheme. In practice, we are interested in the function $f: x \mapsto x^{(P + R)(0)}$. The mapping from $(P + R)(0)$ to $f$ is injective, therefore since $(P + R)(0)$ is a uniformly random value in $\mathbb{Z}_q$, $f$ is a uniformly random function with negligible probability $\frac{1}{q}$ to be sampled.
	
	In order to ensure that $P$ is not of degree $d > k-1$, we compute the DPRF on a pseudo-random input. After each party has published a commitment to their secret key, they are bound to use the same secret key to compute the DPRF. The honest participant generates a random input to the DPRF and requires partial evaluations $\{z_j\}_{j\in Participants}$ from each participant. The honest participant runs $combine(p, q, \{z_i\}_{i\in Participants})$ to use as a reference value. The value is to be compared with $combine(p, q, \{z_i\}_{i\in K})$ for $K\subseteq Participants$ of size exactly $k$, as in line 15 of Algorithm \ref{AlgwithoutTTP}. If $P + R$ was of degree $ d > k-1$, reconstruction would require a subset of $\{z_i\}_{i\in Participants}$ of size $d + 1 > k$\footnote{Note that this means only one test is required as no subset of size $k$ could reconstruct the correct polynomial.}, therefore $combine(p, q, \{z_i\}_{i\in K})$ would output a different value from $combine(p, q, \{z_i\}_{i\in Participants})$. If both values are equal, the honest participant has a guarantee that, with overwhelming probability\footnote{There is a negligible probability that, although the polynomials are different, the value on input $0$ is the same.}, $k$ parties are enough to compute the DPRF.
	
	There is a negligible probability that $P + R$ is of degree $d < k-1$. In order to ensure that this does not happen, the honest participant compares $combine(p, q, \{z_i\}_{i\in Participants})$ with $combine(p, q, \{z_i\}_{i\in D})$ for \\$D\subseteq Participants$ of size exactly $k-1$, as in line 16 of Algorithm \ref{AlgwithoutTTP}. If the values are different, the honest participant has a guarantee that at least $k$ parties are necessary to compute the DPRF.
	
	There is also a negligible probability that $(P + R)(0) = 0$, in which case the DPRF is a constant function equal to the identity of the order $q$ group generated by public parameter $g$. This case can be avoided by asserting that $combine(p, q, \{z_i\}_{i\in Participants})$ is different from the identity, as in line 17 of Algorithm \ref{AlgwithoutTTP}.
	
	At this point, the honest party has the guarantee that they share a $k$-of-$n$ threshold DPRF with the other parties and that the secret function $f$ is uniformly random. Note that the computation of $combine(p, q, \{z_i\}_{i\in Participants})$ and $combine(p, q, \{z_i\}_{i\in K})$ does not reveal $f$ to the honest party but only the output of $f$ for a random input. In other words, it reveals the same information as running an encryption operation. The DPRF is still usable for pseudo-randomly generated inputs, which is how we use it.
\end{proof}

\begin{lemma}
\label{guessing}
	If at most $k-1$ participants are corrupted by an adversary, the adversary can guess the SSSS polynomial with probability at best $\frac{1}{2^\lambda}$.
\end{lemma}

\begin{proof}[Proof of Lemma \ref{guessing}]
	Assume that the adversary has corrupted a subset $T$ of $t = k-1$ participants before setup. The goal of the adversary is to determine the shared polynomial without making the setup fail. This case also proves the point for fewer than $k-1$ corrupted parties as the adversary would hold strictly less information, which can only decrease its ability to determine the shared polynomial.
	
	The adversary knows $k-1$ secret keys $\{sk_i\}_{i\in T}$. Each possible polynomial evaluation value for a $k$-th secret key corresponds to a different polynomial. There are $q$ possible values for this polynomial evaluation, thus, the adversary must determine which of $q$ polynomials is the correct one.
	
	The only additional information known to the adversary about the polynomial are $k-1$ partial secret keys $\{sk_{i,j}\}_{i\in T}$ for each of the remaining $n-k+1$ secret keys $\{sk_j\}_{j\notin T}$. Based on this information, each of these keys $sk_j$ is uniformly random among $q$ possibilities to the adversary according to Statement \ref{Uniformity} (case with $k=1$). Indeed: $$\forall j \notin T \textbf{, } sk_j = \left(\sum_{i\notin T} sk_{i,j}\right) + \left(\sum_{i\in T} sk_{i,j}\right)$$ and $\ds\sum_{i\notin T} sk_{i,j}$ is an unknown uniform value to the adversary.
	
	The adversary is left with the choice to guess either the polynomial with probability $\frac{1}{q}$ or equivalently any of the $n-k+1$ missing shares with probability $\frac{1}{q}$.
\end{proof}
	
	At the end of the setup, the adversary has no additional knowledge on the shared secret compared to $SSNPR$. Breaking pseudo-randomness of our DPRF is therefore has hard as breaking pseudo-randomness of \\$SSNPR$. DiSE \cite{agrawal2018dise} proves that this is at least has hard as solving the DDH problem, which we assume to be hard. This proves the pseudo-randomness of our construction.

	\textbf{Correctness.} When party $i$ requires the help of $K\subseteq Participants$ to compute the DPRF on input $x$, each party $j\in K$ provides an evaluation $z_j$. The proof system used in $SSNPR$ binds each party $j$ to provide $z_j = evaluate(p, q, sk_j, x)$ with $sk_j$ they used in the Pedersen commitment during setup. Since the commitments were published before testing the DPRF, $i$ is guaranteed that the secret keys used to compute $\{z_j\}_{j\in K}$ are the same used during testing. Party $i$ is therefore guaranteed that $\{z_j\}_{j\in K}$ was computed honestly, and that the output of $combine(p, q, \{z_j\}_{j\in K})$ is legitimate.
	
	In order to break correctness, an adversary must break the binding property of the Pedersen commitment. Assuming the hardness of the DDH problem, an adversary can break the binding property of the Pedersen commitment with probability at most $\frac{1}{2^\lambda}$. We refer to DiSE \cite{agrawal2018dise} for a more formal demonstration of the completeness and soundness of the proof system.
\end{proof}

\section{Practical implementation}
\label{implementationChapter}

\subsection{Hashing to an elliptic curve}

	One challenge we faced when implementing the algorithm was the construction of a cryptographic hash function with a Schnorr group as output domain. We use such a hash function twice: to produce an input for the DPRF and to generate a secure generator for the Pedersen commitment scheme. The IETF provides a solution \cite{faz-hernandez_scott_sullivan_wahby_wood_2021} to this challenge that we briefly describe in this subsection.
	
	At line 5 of Encryption and line 4 of Decryption in Algorithm \ref{DiSE}, $i||\alpha$ is not a Schnorr group element, therefore it must be hashed to a Schnorr group element to be used as input to the partial evaluations. However, if a participant knew the discrete logarithm of these inputs relative to a constant generator, then they could compute any output of the DPRF without help after receiving one input. In other words, the DPRF would be secure for only one use.
	
	In the Pedersen commitment scheme, we need two generators such that the logarithm of the second relative to the first is unknown. If the committer knew that logarithm, they could break the binding property of the commitment. In practice, the generators are often generated by a TTP, which we do not have available in this implementation. In order to avoid this issue, we always use the same first generator, while we generate a random second generator of the Schnorr group for every instance of the scheme. We generate this element by hashing a random seed to an element of the Schnorr group and making both public. Since Schnorr groups are of prime order, any element except identity is a generator.\\
	
	A solution to hashing to a Schnorr group without revealing the discrete logarithm of the output is to use elliptic curves. Detailed hash functions onto elliptic curves can be found in the IETF \cite{faz-hernandez_scott_sullivan_wahby_wood_2021}. In our case, we are interested in hashing to secp256k1. In order to implement the solution, we use 3 primitives\footnote{Construction of these primitives is explained in \cite{faz-hernandez_scott_sullivan_wahby_wood_2021}}:
\begin{enumerate}
	\item an extendable output function $XOF$ such that $XOF(m, n)$ outputs $n$ values in $\mathbb{Z}_q$
	\item a mapping $M$ to a curve $\mathcal{C}$ isogenous to secp256k1
	\item the isogeny $E$ from $\mathcal{C}$ to secp256k1
\end{enumerate}
	From these, we construct the hash function to secp256k1 as in Algorithm \ref{hashingAlg}.

\subsection{Results}

	We implement our scheme in Python and run simulations with each entity in its own process to simulate a realistic environment. The commitment scheme used in line 3 of encryption in Algorithm \ref{DiSE} is a hash-based commitment using shake256. The DDH-based DPRF as described in \cite{naor1999distributed} is constructed on secp256k1. We use AES counter mode for the PRNG primitive.

	We also implement DiSE with the same primitives and run simulations on the same machine for comparison. The timing results obtained aim at comparing the speed of both algorithm. However, an optimized implementation would run significantly faster.\\
	
	After completing setup, a participant must store the following values:\begin{enumerate}
	\item a 64 bytes secret share for the DPRF
	\item 32 bytes of randomness used for proving partial evaluation
	\item a 16 bytes personal identifier
	\item a 32 bytes private key for authentication
\end{enumerate}
In addition, for each of the $n$ participants in the scheme, one participant needs to store an identifier, an address and a public key, which adds up to 86 bytes. This data could alternatively be stored in a semi-trusted database to alleviate the storage cost on the participants' devices. This does not include the storage cost of an additional whitelist to determine rights to request operations for specific participants.\\
	
	Table \ref{setupTiming} was obtained by timing the execution of 100 setups with various number of participants $n$ and threshold $k$ values. The setups are launched sequentially.

\begin{table}[h]
\caption{Setup performance metrics}
\label{setupTiming}
\centering
\begin{tabular}{|c|c||c|c||c|c|}
	\hline
	$k$ & $n$ & \multicolumn{2}{c||}{Throughput} & \multicolumn{2}{c|}{Latency} \\
	  &  & \multicolumn{2}{c||}{(\emph{setup/s})} & \multicolumn{2}{c|}{(\emph{ms/setup})} \\
	\cline{3-6}
	  &  & TDiSE & DiSE & TDiSE & DiSE \\
	\hline
	$n/3$ & 6 & 8.30 & 11.8 & 7.23 & 5.08 \\
	\cline{2-6}
	  & 12 & 4.20 & 11.6 & 14.3 & 5.15 \\
	\cline{2-6}
	  & 18 & 2.31 & 11.5 & 25.9 & 5.22 \\
	\cline{2-6}
	  & 24 & 1.43 & 11.3 & 42.1 & 5.29 \\
	\hline
	\hline
	$n/2$ & 4 & 9.90 & 11.9 & 6.06 & 5.06 \\
	\cline{2-6}
	  & 6 & 8.07 & 11.8 & 7.44 & 5.08 \\
	\cline{2-6}
	  & 12 & 4.00 & 11.6 & 15.0 & 5.16 \\
	\cline{2-6}
	  & 18 & 2.15 & 11.5 & 28.0 & 5.22 \\
	\cline{2-6}
	  & 24 & 1.34 & 11.3 & 44.7 & 5.29 \\
	\hline
	\hline
	$2n/3$ & 3 & 10.6 & 11.9 & 5.64 & 5.05 \\
	\cline{2-6}
	  & 6 & 7.91 & 11.8 & 7.58 & 5.09 \\
	\cline{2-6}
	  & 12 & 3.83 & 11.6  & 15.7 & 5.15\\
	\cline{2-6}
	  & 18 & 2.05 & 11.5 & 29.3 & 5.22 \\
	\cline{2-6}
	  & 24 & 1.25 & 11.3 & 47.9 & 5.29 \\
	\hline
	\hline
	$3n/4$ & 4 & 9.67 & 11.8 & 6.21 & 5.06 \\
	\cline{2-6}
	  & 12 & 3.74 & 11.6 & 16.1 & 5.16 \\
	\cline{2-6}
	  & 24 & 1.20 & 11.3 & 50.1 & 5.30 \\
	\hline
	\hline
	$n-2$ & 12 & 3.59 & 11.6 & 16.7 & 5.15 \\
	\cline{2-6}
	  & 18 & 1.89 & 11.5 & 31.7 & 5.22 \\
	\cline{2-6}
	  & 24 & 1.15 & 11.3 & 52.0 & 5.29 \\
	\hline
	\hline
	$2$ & 12 & 4.52 & 11.6 & 13.3 & 5.15 \\
	\cline{2-6}
	  & 18 & 2.53 & 11.5 & 23.7 & 5.22 \\
	\cline{2-6}
	  & 24 & 1.60 & 11.3 & 37.5 & 5.29 \\
	\hline
\end{tabular}
\end{table}

	The setup operation of TDiSE requires $\theta(n^2)$ communications as compared to $n$ communications for DiSE. These communication complexity estimations are supported by the results in Table \ref{setupTiming}. Compared to DiSE, the setup of TDiSE pays a heavier toll when increasing the $n$ parameter. We can also observe that for constant $n$ values, the setup duration increases with $k$. This impact remains however negligible compared to that of $n$. \\
	
	Table \ref{encTiming} was obtained by timing the encryption of 100 messages of 32 bytes each with various number of participants $n$ and threshold $k$ values. The encryptions are run sequentially.

\begin{table}[h]
\caption{Encryption performance metrics}
\label{encTiming}
\centering
\resizebox{7.5cm}{5cm}{
\begin{tabular}{|c|c||c|c||c|c||c|c|}
	\hline
	$k$ & $n$ & \multicolumn{2}{c||}{Throughput} & \multicolumn{2}{c||}{Latency} & \multicolumn{2}{c|}{Bandwidth} \\
	  &  & \multicolumn{2}{c||}{(\emph{enc/s})} & \multicolumn{2}{c||}{(\emph{ms/enc})} & \multicolumn{2}{c|}{(Throughput \emph{Kbps})} \\
	\cline{3-8}
	  &  & TDiSE & DiSE & TDiSE & DiSE & TDiSE & DiSE \\
	\hline
	$n/3$ & 6 & 11.7 & 11.6 & 85.7 & 86.1 & 2.99 & 2.97 \\
	\cline{2-8}
	  & 12 & 5.69 & 5.83 & 176 & 171 & 1.46 & 1.49 \\
	\cline{2-8}
	  & 18 & 3.78 & 3.84 & 264 & 260 & 0.97 & 0.98 \\
	\cline{2-8}
	  & 24 & 2.82 & 2.84 & 355 & 352 & 0.72 & 0.73 \\
	\hline
	\hline
	$n/2$ & 4 & 11.5 & 11.6 & 86.9 & 86.5 & 2.95 & 3.00 \\
	\cline{2-8}
	  & 6 & 7.61 & 7.79 & 131 & 128 & 1.95 & 1.99 \\
	\cline{2-8}
	  & 12 & 3.71 & 3.86 & 270 & 259 & 0.95 & 0.99 \\
	\cline{2-8}
	  & 18 & 2.55 & 2.52 & 392 & 397 & 0.65 & 0.64 \\
	\cline{2-8}
	  & 24 & 1.88 & 1.91 & 531 & 522 & 0.48 & 0.49 \\
	\hline
	\hline
	$2n/3$ & 3 & 11.7 & 11.6 & 85.6 & 86.2 & 2.99 & 2.97 \\
	\cline{2-8}
	  & 6 & 5.94 & 5.82 & 168 & 172 & 1.52 & 1.49 \\
	\cline{2-8}
	  & 12 & 2.91 & 2.90 & 344 & 345 & 0.75 & 0.74 \\
	\cline{2-8}
	  & 18 & 1.90 & 1.87 & 525 & 535 & 0.49 & 0.48 \\
	\cline{2-8}
	  & 24 & 1.41 & 1.41 & 707 & 707 & 0.36 & 0.36 \\
	\hline
	\hline
	$3n/4$ & 4 & 7.18 & 7.58 & 139 & 132 & 1.84 & 1.94 \\
	\cline{2-8}
	  & 12 & 2.45 & 2.52 & 408 & 397 & 0.63 & 0.64 \\
	\cline{2-8}
	  & 24 & 1.24 & 1.23 & 807 & 809 & 0.32 & 0.32 \\
	\hline
	\hline
	$n-2$ & 12 & 2.29 & 2.28 & 437 & 439 & 0.59 & 0.58 \\
	\cline{2-8}
	  & 18 & 1.41 & 1.42 & 709 & 706 & 0.36 & 0.36 \\
	\cline{2-8}
	  & 24 & 1.02 & 1.03 & 976 & 970 & 0.26 & 0.26 \\
	\hline
	\hline
	$2$ & 12 & 11.5 & 11.5 & 86.7 & 86.7 & 2.95 & 2.95 \\
	\cline{2-8}
	  & 18 & 11.6 & 11.4 & 86.1 & 87.6 & 2.97 & 2.92 \\
	\cline{2-8}
	  & 24 & 11.4 & 11.4 & 87.7 & 87.8 & 2.92 & 2.92 \\
	\hline
\end{tabular}
}
\end{table}

	The encryption operation requires $\theta(k)$ communications for both algorithms. These communications are independent from one another, therefore they could be run in parallel. We chose to run the communications sequentially for testing, resulting in an upper bound of timing. As expected, Table \ref{encTiming} shows that both algorithms perform at the same speed and depend only on the $k$ parameter.

\section{Proactive security}
\label{proactiveChapter}

	It is possible to upgrade our protocol with proactive security by periodically updating the secret keys without changing the shared secret. This would make an adversary's task more challenging, since they must now compromise at least $k$ keys before they are updated.
	
	In order to renew keys, the untrusted third party prompts each participant to run Algorithm \ref{updatekeys}. Each participant generates a set of SSSS keys for a polynomial of constant value $0$ and distributes the keys to each participant with the same attribution convention as our setup. By summing all received keys, a participant obtains a new secret key that can be added to their previous key without changing the DPRF. This way the DPRF remains unchanged while making older keys obsolete. Analogously to Algorithm \ref{AlgwithoutTTP}, the untrusted third party prompts participants to test their new keys to ensure that they share a $k$-of-$n$ threshold DPRF. An additional test is added to ensure that the output of the DPRF is consistent with the ouput obtained with the previous set of keys.
	
\begin{algorithm}[h]
\caption{$UpdateKeys_{TDiSE}$ run by participant $i$}
\label{updatekeys}
\fontsize{8pt}{8pt}\selectfont
\begin{algorithmic}[1]
	\Require $p(\lambda), q(\lambda), k, n, \{x_j\}_{j\in \llbracket n \rrbracket}, Participants$
	\Ensure $p, q$
	
	\hspace{-4em} \textbf{Private output:} $SK'$
	\State Run $generate_{SSSS}(q(\lambda), k, n, \{x_j\}_{j\in \llbracket n \rrbracket}, 0, Participants)$ to distribute new secret keys $(sk'_{i,1}, ..., sk'_{i,n})$
	\For{$j\in Participants$}
		\State Receive $sk'_{j,i}$ from participant $j$
	\EndFor
	\State Compute $sk'_i = sk_i +  \sum\limits_{j\in\llbracket n\rrbracket} sk'_{j, i}$
	\State Publish $\gamma '_i = commit_{Pedersen}(sk'_i)$
	\State Wait for every $j\in Participants$ to publish a commitment
	\State Generate randomness $\rho$
	\State Generate message $m = H_{test}(p, q, \lambda, k, n, Participants, \{\gamma '_j\}_{j\in Participants})$
	\State Compute commitment $\alpha = commit_{hash-based}(m, \rho)$
	\For{ $j \in Participants$  }
		\State Send $i || \alpha$ to participant $j$
		\State Obtain $z_j = evaluate_{SSNPR}(p, q, sk_j, i || \alpha)$ from participant $j$
		\State Obtain $z'_j = evaluate_{SSNPR}(p, q, sk'_j, i || \alpha)$ from participant $j$
	\EndFor
	\State Assert that 
	
	\hspace{-2em} $combine_{SSNPR}(p, q, \{z_{i,j}\}_{j\in Participants}) = combine_{SSNPR}(p, q, \{z'_{i,j}\}_{j\in Participants})$
	\State Assert that for $K\subset Participants$ of size exactly $k$, 
	
	\hspace{-2em} $combine_{SSNPR}(p, q, \{z'_{i,j}\}_{j\in K}) = combine_{SSNPR}(p, q, \{z'_{i,j}\}_{j\in Participants})$
	\State Assert that for $T\subset Participants$ of size exactly $k-1$, 
	
	\hspace{-2em} $combine_{SSNPR}(p, q, \{z'_{i,j}\}_{j\in T}) \neq combine_{SSNPR}(p, q, \{z'_{i,j}\}_{j\in Participants})$
	\State Assert that $combine_{SSNPR}(p, q, \{z'_{i,j}\}_{j\in Participants}) \neq 1$
	\State \textbf{Return} $p, q$
\end{algorithmic}
\end{algorithm}
	
	The security proof of this key update protocol is the same as the security proof of the setup. The only difference is that an additional test must be added to the protocol to ensure that the DPRF still outputs the same values as it would have with the previous keys.
	
\section{Conclusion}
\label{conclusion}

	In this work, we have presented a TSE scheme that does not require to trust a party to ever hold all the keys at once. Our scheme facilitates the distribution of encryption capacity among entities without requiring them to trust each other. We achieve this by modifying the setup of DiSE \cite{agrawal2018dise}, such that, instead of using a TTP to generate and distribute the keys, each party contributes to the key generation in a way that only reveals each partial key to its recipient party. The security of our scheme holds in the Random Oracle Model assuming the hardness of the DDH problem. Removing the TTP comes with an additional cost in complexity for the setup protocol, however the setup remains of polynomial complexity and its cost is amortized as one setup execution is sufficient for several encryption and decryption rounds. \\
	
	A formal proof of security in the model defined by ParaDiSE \cite{agrawal2022paradise} has yet to be written, which we leave to future works.
	
	Since our implementation is based on the DDH assumption, its security is compromised in the post-quantum context. An axis of improvement is to develop a quantum-resistant dealer-free threshold symmetric encryption scheme by replacing vulnerable primitives in our construction. Similarly to the work of \cite{sinha2024efficient}, the most promising avenue is replacing our DPRF with a verifiable DPRF based on the LWE problem.

\subsubsection*{Acknowledgements} This research received the support of Mitacs in the context of the Mitacs Accélération program.

\bibliographystyle{splncs04}
\bibliography{MyBib}

\appendix
\section*{Appendix: Algorithms}

\begin{algorithm}
\caption{Hash function $H$ to secp256k1}
\label{hashingAlg}
\fontsize{8pt}{8pt}\selectfont
\begin{algorithmic}[1]
	\Require message $m$
	\Ensure a point $P$ on secp256k1
	\State Compute $u,v = XOF(m, 2)$
	\State Compute $P0 = E(M(u))$
	\State Compute $P1 = E(M(v))$
	\State Compute $P = P0 + P1$
	\State \textbf{Return} $P$
\end{algorithmic}
\end{algorithm}

\begin{algorithm}
\caption{Shamir's Secret-Sharing Scheme}
\label{SSSS}
\fontsize{8pt}{8pt}\selectfont
\begin{algorithmic}[1]
	\Algphase{$generate_{SSSS}$}
	\Require $q(\lambda), k, n, \{x_j\}_{j\in\llbracket n\rrbracket}, s,  Participants$
	\Ensure $q, k, n, \{x_j\}_{j\in\llbracket n\rrbracket}$
	
	\hspace{-4em} \textbf{Private output:} $SK$
	\State Sample $a_1, ..., a_{k-1}$ uniformly in $\mathbb{Z}_q$
	\State Define $\ds P(X) = s + \sum_{i=1}^{k-1} a_i X^i$
	\State Set $SK = \{sk_j =(x_j, P(x_j) \text{ mod } q) \text{ for } j\in\llbracket n\rrbracket\}$
	\For{ $j \in \llbracket n\rrbracket$}
		\State Send $sk_j\in SK$ to participant $j$ in $Participants$
	\EndFor
	\State \textbf{Return} $q, k, n, \{x_j\}_{j\in\llbracket n\rrbracket}$
\end{algorithmic}
\begin{algorithmic}[1]
	\Algphase{$combine_{SSSS}$}
	\Require $q, k, n, \{sk_j\}_{j\in K}$
	\Ensure $s$
	\State Parse the secret keys as $\{sk_j\}_{j\in K} = \{(x_j, P(x_j))\}_{j\in K}$
	\For{ $j \in K$ }
		\State Compute $\lambda_j = \prod_{i\in K, i \neq j} \frac{x_i}{x_i-x_j}$
	\EndFor
	\State \textbf{Return} $\ds\sum_{j\in K} \lambda_j P(x_j) $
\end{algorithmic}
\end{algorithm}

\begin{algorithm}
\caption{$SSNPR$}
\label{SSNPR}
\fontsize{8pt}{8pt}\selectfont
\begin{algorithmic}[1]
	\Algphase{$generate_{SSNPR}$}
	\Require $p(\lambda), q(\lambda), k, n, \{x_j\}_{j\in\llbracket n\rrbracket}, Participants$
	\Ensure $p, q, k, n, \{x_j\}_{j\in\llbracket n\rrbracket}, g, h, \{\gamma_j\}_{j\in \llbracket n\rrbracket}$
	
	\hspace{-4em} \textbf{Private output:} $SK, \{r_j\}_{j\in Participants}$
	\State Sample $s$ uniformly in $\mathbb{Z}_q$
	\State Run $generate_{SSSS}(q(\lambda), k, n, \{x_j\}_{j\in\llbracket n\rrbracket}, s, Participants)$
	\State Run $setup_{Pedersen}(\lambda)$ to obtain $g$ and $h$
	\For{ $j \in \llbracket n\rrbracket$}
		\State Sample $r_j$ uniformly in $\mathbb{Z}_q$
		\State Compute $\gamma_j = commit_{Pedersen}(p, g, h, sk_j, r_j)$
		\State Send $r_j$ to participant $j$ in Participants
	\EndFor
	\State \textbf{Return} $p, q, k, n, \{x_j\}_{j\in\llbracket n\rrbracket}, g, h, \{\gamma_j\}_{j\in \llbracket n\rrbracket}$
\end{algorithmic}
\begin{algorithmic}[1]
	\Algphase{$evaluate_{SSNPR}$}
	\Require $p, q, g, h, \{\gamma_j\}_{j\in \llbracket n\rrbracket}, sk_i, x$
	\Ensure $z_i$
	\State Compute $\omega = H(x)$
	\State Compute $h_i =\omega^{sk_i}$
	\State Sample $v_i$ and $v_i'$ uniformly in $\mathbb{Z}_q$
	\State Compute $t_i = \omega^{v_i}$ 
	\State Compute $t_i' = g^{v_i}. h^{v'i}$
	\State Compute $c_i = H_{chal}(h_i, \omega, \gamma_i, g, h, t_i, t_i')$
	\State Compute $u_i = v_i - c_i\cdot sk_i$
	\State Compute $u_i' = v_i' - c_i\cdot r_i$
	\State \textbf{Return} $z_i = (h_i, c_i, u_i, u_i')$
\end{algorithmic}
\begin{algorithmic}[1]
	\Algphase{$combine_{SSNPR}$}
	\Require $p, q, \{x_j\}_{j\in\llbracket n\rrbracket}, g, h, \{\gamma_j\}_{j\in \llbracket n\rrbracket}, \{z_j\}_{j\in K}$
	\Ensure $\omega^s$
	\For{ $j\in K$ }
		\State Parse $z_j = h_j, c_j, u_j, u_j'$
		\State Compute $t_j = \omega^{u_j}\cdot h_j^{c_j}$
		\State Compute $t_j' = g^{u_j}\cdot h^{u_j'}\cdot\gamma_j^{c_j}$
		\State Assert that $c_j = H_{chal}(h_i, \omega, \gamma_j, g, h, t_j, t_j')$
	\EndFor
	\For{ $j \in K$ }
		\State Compute $\lambda_j = \prod_{i\in K, i \neq j} \frac{x_i}{x_i-x_j}$
	\EndFor
	\State \textbf{Return} $\omega^s = \ds\prod_{j\in K} z_j^{\lambda_j}$
\end{algorithmic}
\end{algorithm}

\begin{algorithm}
\caption{$NPR$}
\label{NPR}
\fontsize{8pt}{8pt}\selectfont
\begin{algorithmic}[1]
	\Algphase{$generate_{NPR}$}
	\Require $p(\lambda), q(\lambda), k, n, \{x_j\}_{j\in\llbracket n\rrbracket}, Participants$
	\Ensure $p, q, k, n, \{x_j\}_{j\in\llbracket n\rrbracket}$
	
	\hspace{-4em} \textbf{Private output:} $SK$
	\State Sample $s$ uniformly in $\mathbb{Z}_q$
	\State Run $generate_{SSSS}(q(\lambda), k, n, \{x_j\}_{j\in\llbracket n\rrbracket}, s, Participants)$
	\State \textbf{Return} $p, q, k, n, \{x_j\}_{j\in\llbracket n\rrbracket}$
\end{algorithmic}
\begin{algorithmic}[1]
	\Algphase{$evaluate_{NPR}$}
	\Require $p, sk_i, x$
	\Ensure $z_i$
	\State Compute $\omega = H(x)$
	\State \textbf{Return} $z_i = \omega^{sk_i}$
\end{algorithmic}
\begin{algorithmic}[1]
	\Algphase{$combine_{NPR}$}
	\Require $p, q, \{x_j\}_{j\in\llbracket n\rrbracket}, \{z_j\}_{j\in K}$
	\Ensure $\omega^s$
	\For{ $j \in K$ }
		\State Compute $\lambda_j = \prod_{i\in K, i \neq j} \frac{x_i}{x_i-x_j}$
	\EndFor
	\State \textbf{Return} $\omega^s = \ds\prod_{j\in K} z_j^{\lambda_j}$
\end{algorithmic}
\end{algorithm}

\begin{algorithm}
\caption{DiSE}
\label{DiSE}
\fontsize{8pt}{8pt}\selectfont
\begin{algorithmic}[1]
	\Algphase{Setup}
	\Require $\lambda, k, n$
	\Ensure public parameters $pp_{TSE}$
	
	\hspace{-4em} \textbf{Private output:} $SK$
	\State Run $generate_\DPRF(\lambda, k, n)$ to obtain $pp_\DPRF$ and distribute $SK$
	\State Run $setup_\Comm(\lambda)$ to obtain $pp_\Comm$
	\State \textbf{Return} $pp_{TSE} = (pp_\DPRF, pp_\Comm)$
\end{algorithmic}
\begin{algorithmic}[1]
	\Algphase{Encryption by participant $i$}
	\Require $pp_{TSE}$, message $m$, $K\subseteq Participants$
	\Ensure ciphertext $c$
	\State Parse public parameters as $pp_{TSE} = (pp_\DPRF, pp_\Comm)$
	\State Generate randomness $\rho$
	\State Compute commitment $\alpha = commit_\Comm(pp_\Comm, m, \rho)$
	\For{ $j \in K$  }
		\State Send $i || \alpha$ to participant $j$
		\State Obtain $z_j = evaluate_\DPRF(pp_\DPRF, sk_j, i || \alpha)$ from participant $j$
	\EndFor
	\State Compute $\omega = combine_\DPRF(pp_\DPRF, \{z_j\}_{j \in K})$
	\State Compute $\epsilon = PRNG(\omega)\oplus (m || \rho)$
	\State \textbf{Return} $c = (i, \alpha, \epsilon)$
\end{algorithmic}
\begin{algorithmic}[1]
	\Algphase{Decryption by participant $i$}
	\Require $pp_{TSE}$, ciphertext $c$, $K\subseteq Participants$
	\Ensure message $m$
	\State Parse public parameters as $pp_{TSE} = (pp_\DPRF, pp_\Comm)$
	\State Parse ciphertext as $c = (i, \alpha, \epsilon)$
	\For{ $j \in K$  }
		\State Send $i || \alpha$ to participant $j$
		\State Obtain $z_j = evaluate_\DPRF(pp_\DPRF, sk_j, i || \alpha)$ from participant $j$
	\EndFor
	\State Compute $\omega = combine_\DPRF(pp_\DPRF, \{z_j\}_{j \in K})$
	\State Recover $m || \rho = PRNG(\omega)\oplus \epsilon$
	\If{ $verify_\Comm(pp_\Comm, m, \rho, \alpha) \neq true$}
		\State \textbf{Return} failure
	\EndIf
	\State \textbf{Return} $m$
\end{algorithmic}
\end{algorithm}

\end{document}